\documentstyle[prl,aps, epsf,preprint]{revtex}
 
\renewcommand{\vec}[1]{{\mathbf #1}}

\begin{document}  
\title{ Dynamics of Interacting Neural Networks}   
\author{W. Kinzel and R. Metzler}
\address{Institut f\"{u}r Theoretische Physik, 
Universit\"{a}t W\"{u}rzburg, Am Hubland, D-97074 W\"{u}rzburg, Germany}
\author{I. Kanter}
\address{Minerva Center and Department of Physics, Bar Ilan University, 52900 
Ramat Gan, Israel}  
\date{\today}
\maketitle
\begin{abstract}  
  The dynamics of interacting perceptrons is solved analytically.  For a
  directed flow of information the system runs into a state which has a
  higher symmetry than the topology of the model. A symmetry breaking
  phase transition is found with increasing learning rate.  In addition
  it is shown that a system of interacting perceptrons which is trained
  on the history of its minority decisions 
  develops a good strategy for the problem of adaptive competition known
  as the Bar Problem or Minority Game.
\end{abstract}

Simple models of neural networks describe a wide variety of phenomena in
neurobiology and information theory. Neural networks are systems of
elements interacting by adaptive couplings which are trained by a set of
examples. After training they function as content addressable
associative memory, as classifiers or as prediction algorithms. Using
methods of statistical physics many of these phenomena have been
elucidated analytically for infinitely large neural networks
\cite{Hertz,Opper}.

Up to now, only isolated neural networks have been investigated
theoretically. However, many phenomena in biology, social science and
computer science may be modelled by a system of interacting adaptive
algorithms.  Nothing is known about general properties of such systems.
In this Letter we present the first analytic solution of a system of
interacting perceptrons. For simplicity, we restrict ourselves to simple
perceptrons with binary output.

The dynamics of a set of perceptrons learning from each other by a
directed flow of information is solved analytically.  Starting from a
nonsymmetric initial configuration, the system relaxes to a final state
which has a higher symmetry than the ring like flow of information. The
system tries to stay as symmetric as possible.  In some cases we find a
phase transition: When the learning rate is increased the system
suddenly breaks the symmetry and relaxes to a state with nonsymmetric
overlaps.

In addition we show that a system of interacting neural networks can
develop a good strategy for a model of adaptive competition in closed
markets, the El-Farol Bar problem \cite{Arthur}.
In this problem $N$ agents are competing for limited resources and the
individual profit depends on the collective behaviour. Recently a
variation of this problem known as the Minority Game has been 
studied theoretically in a series of publications
\cite{Challet,Savit,Cavagna,Challet:Exact,Challet:Modeling}. 
The Minority Game model has several 
peculiarities: (a) The strategies of the agents are
quenched random variables (decision tables), given in advance, 
and each agent can only choose between a few of these tables. 
(b) Some of the agents are
frozen as losers, at least in some regions of the parameter space. 
In a realistic situation permanent losers would change the strategy
after some time. (c) A good performance is only achieved 
if the number of time steps each agent is using for his/her 
decison is adjusted to the number of agents.

Our approach shows none of these drawbacks. Each agent is using one
perceptron for his/her decision with couplings which are trained to the
minority of all the outputs. Hence the strategies develop according to
the dynamics of the system.  We analytically calculate the statistical
properties of such a system of interacting perceptrons. The system
performs optimally in the limit of small learning rates and is
insensitive to the size of the number of time steps taken for the
decision.  Each agent receives the same profit in the long run.
  
The perceptron is the simplest model of a neural network. It has one layer
of synaptic weights $\vec{w} = (w_1, ..., w_M)$ and one output bit
$\sigma$ which is given by 
\begin{equation} \label{eins} \sigma = \mbox{sign}
\sum\limits^{M}_{i=1} \; w_i x_i = \mbox{sign} (\vec{w} \cdot
\vec{x}) \;  \; . \end{equation} 
$\vec{x}$ is the input vector
of dimension $M$; for instance it is given by a window of a bit sequence
$S_t \in \{+1,-1\}, t=1, 2, \dots , M,$ with $\vec{x}_i = ( S_{t -M+1}, 
, \dots, S_{t})$, or it consists of random binary or
gaussian variables. A training example is a pair of input vector and
output bit $(\vec{x} , \sigma)$; a perceptron learns this example by
adjusting its weights to it. Here we consider three well-known learning
rules \cite{Hertz}:\\ 
$H$: Hebbian learning, 
\begin{equation}
\label{zwei} \vec{w}_{\mbox{\footnotesize new }} = 
\vec{w}_{\mbox{\footnotesize old }} +
\frac{\eta}{M} \sigma\, \vec{x} \; \; . \label{Hebb-Eq}
\end{equation} 
$P$:
Perceptron learning: $H$ is applied only if the example is misclassified,
$\vec{w}_{\mbox{\footnotesize old }} \cdot \vec{x}\; \sigma < 0$.\\
$PN$: Learning with normalization: After each step $P$ the
weights are normalized, $\vec{w}_{\mbox{\footnotesize new}} \cdot
\vec{w}_{\mbox{\footnotesize new}} = 1$.

$\eta$ is the size of the learning step. In the following we mainly
consider the limit of infinitely large networks, $M \rightarrow \infty$,
in which the learning step $\eta$ becomes a learning rate for a continuous
presentation of examples. In this case we can use  the analytical methods for
on-line training which are well developed \cite{Opper,Biehl,Saad}.
 In this Letter we study a system of $N$
perceptrons with weight vectors $\vec{w}^1, \vec{w}^2, \ldots \vec{w}^N$
which are trained by a common input vector $\vec{x}$ and their mutual
output bits $\sigma^1 , \ldots , \sigma^N$.

%

\begin{figure} 
\epsfxsize= 0.49\textwidth
 \epsffile{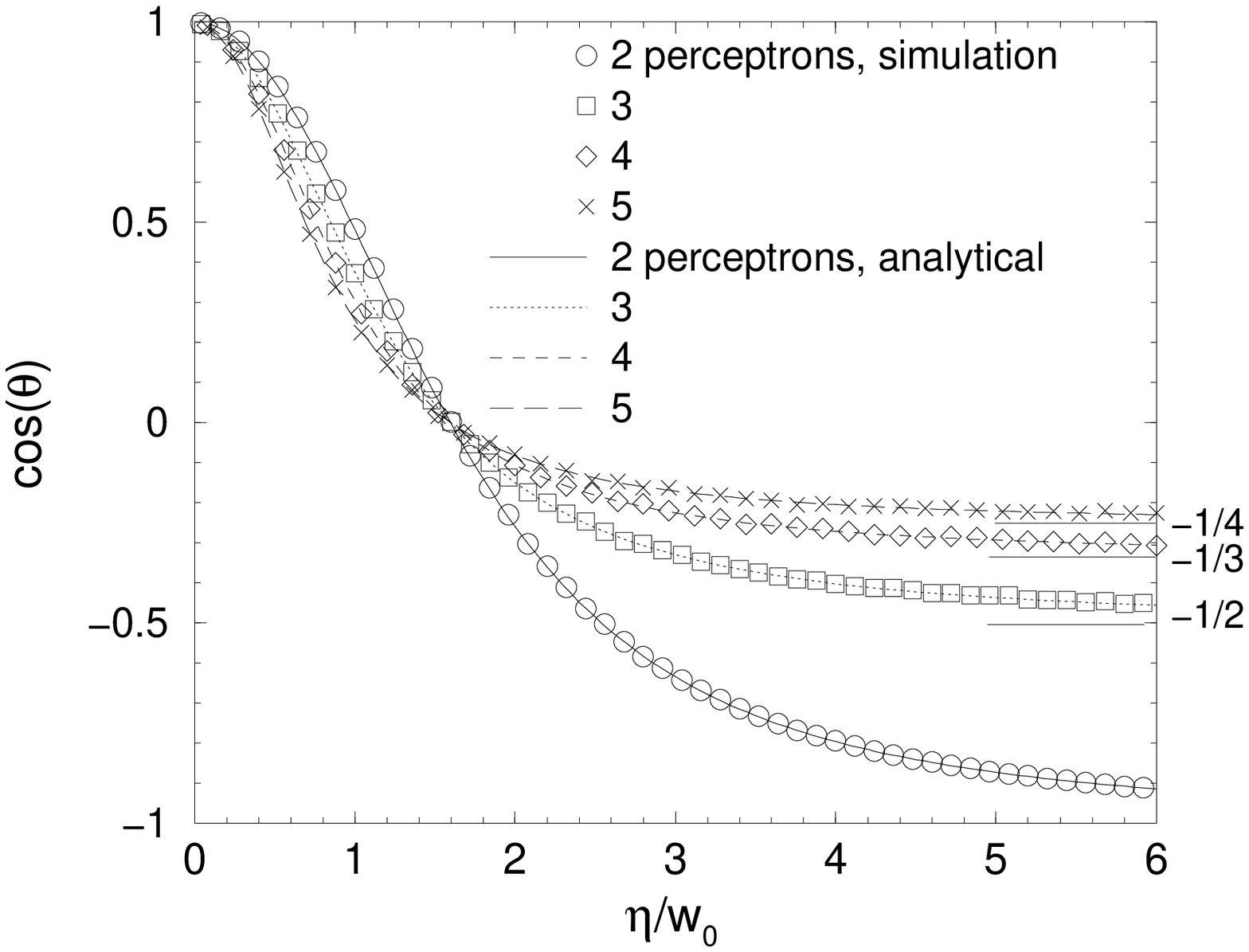}  
\caption{Fixed point of cyclic learning with alg. $P$:
simulations with $M=100$ for 2 to 5 perceptrons and
solutions of Eqn. \ref{fuenf}. $\theta$ is the common 
mutaual angle between all weight vectors.}
\label{GSL-P}
\end{figure}

We consider a set of $N$ interacting perceptrons with a directed cyclic
flow of information. At each training step all of the networks receive the
same randomly chosen input vector $\vec{x}$. Now perceptron
$\vec{w}^1$ learn the output from $\vec{w}^2$, perceptron
$\vec{w}^2$ learns from $\vec{w}^3$ , $\dots$, perceptron
$\vec{w}^N$ learns from perceptron $\vec{w}^1$. Our analytical
and numerical calculations give the following result:  Starting from
random initial weight vectors with length $w_0 = |\vec{w}_0| $ 
and using perceptron learning rule $P$ for each of
the networks, the system runs into a state of complete symmetry with
identical overlaps $\vec{w}^i \cdot \vec{w}^j$ for all pairs
$(i, j)$. The stationary state is given by the equation 
\begin{equation}
\label{fuenf} 
\eta \; \theta \sqrt{1 + (N-1) \cos \; \theta} = \sqrt{2\pi}
\; w_0 \; (1 - \cos \; \theta)\;\; , 
\end{equation} 
where $\theta$ is the common mutual angle between all weight
vectors. Fig. \ref{GSL-P} shows the
result. For small learning rate all perceptrons agree with each other,
their mutual angle $\theta$ is close to zero. 
With increasing learning rate the
angle increases to its maximal possible value. 
The sum of the weight vectors $\sum^N_{i =1} \; \vec{w}^i$ is constant
under this learning rule, because for every perceptron that learns
the pattern with a positive sign there is a subsequent neighbour that learns 
it with a negative sign.
For $ \eta \rightarrow
\infty$ the norm of this sum is negligible compared
to $|\vec{w}^i|$, and the vectors form a hypertetrahedron which gives
$\cos \;\theta = -1/(N-1)$. Note that the final stationary state has a
higher symmetry than the ring flow of information. This symmetry
seems to be robust to details of the model: in simulations where
each perceptron had a different quenched learning rate, all 
the angles between the perceptrons again converged to the same
value.

The effective repulsion between the weight vectors can be
understood geometrically in the case of two perceptrons: the sum 
$\vec{w}^1 + \vec{w}^2$ is conserved; the fixed point
results from an equilibrium between learning the 
component of $\vec{x}$ parallel to the $\vec{w}^1$--$\vec{w}^2$--plane
(which decreases $\theta$) and the component perpendicular
to this plane (which increases $\theta$).

\begin{figure} 
\epsfxsize= 0.49\textwidth
\epsffile{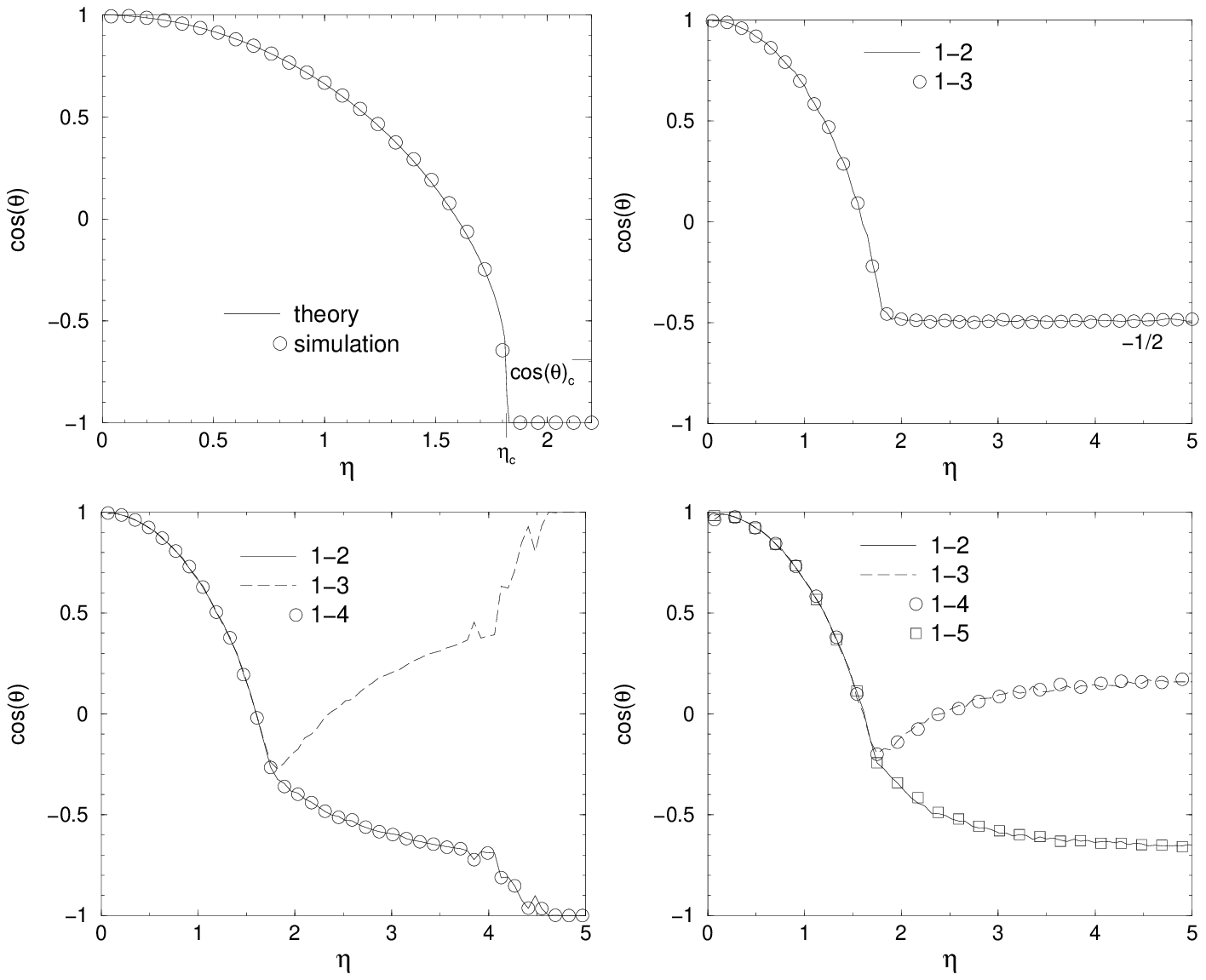}  
\caption{Fixed points of $\cos \theta$ in cyclic 
learning using rule $PN$ for 2, 3, 4, and 5 perceptrons, 
respectively, in simulations with $M=100$.
For $\eta<\eta_c$ all weight vectors have a common mutual
angle $\theta$. For $\eta > \eta_{c}(N)$ and a ring with more
than 3 perceptrons, the symmetry is broken, and the
angle $\theta_{ij}$ depends on the distance between 
perceptrons $i$ and $j$ on the ring.}  
\label{GSL-PN}
\end{figure}

The symmetric behaviour turns out to be different with learning rule $PN$, 
where all the weight
vectors $\vec{w}^i$ remain on a sphere $|\vec{w}^i| = 1$. For
small learning rate the system runs into a symmetric state 
given by 
\begin{equation} \label{fuenfB}
\eta\, \theta = \sqrt{2 \pi} (1- \cos \theta)
\end{equation} 
(compare to Eqn. \ref{fuenf}).
However, this equation can only be geometrically realized up 
to a critical value $\eta_c (N)$, where the hypertetrahedron 
configuration is reached and the sum of the $\vec{w}^i$
vanishes. In the case of two perceptrons, geometrical 
constraints do not play a role; however, there is a maximal 
$\eta_{c,2} \doteq 1.82$, above which no solution of Eqn. \ref{fuenfB}
exists.
For larger learning rates $\eta > \eta_c(N)$ our numerical simulations 
give the
following results as shown in Fig. \ref{GSL-PN}:
\begin{itemize} 
\item For $N=2$, there is a discontinuous transition
to $\cos \theta =-1$ at the mentioned $\eta_{c,2}$.
\item
 For $N=3$ the state remains in the triangular configuration 
$\cos \; \theta = - 1/2$.
\item 
For $N \ge 3$ the symmetry is broken spontaneously. The angle 
$\theta_{ij}$ between perceptrons $i$ and $j$ now depends on
their distance on the ring. 
However, the symmetry of the ring is still conserved. This means, for
instance, that for $N=7$, $\theta_{13}$ is the same as $\theta_{24}$ and
$\theta_{35}$, but there are three different values of mutual angles
$\theta_{12}, \theta_{13}$ and $\theta_{14}$. In general, 
there are now $N/2$ different angles for even $N$ and $(N-1)/2$
angles for odd $N$. Since the perceptrons try to increase
the angle to their nearest neighbour, the angle to more
distant perceptrons has to increase to satisfy geometric
constraints.
\item 
For even values $N\ge 4$ we observe an additional discontinuous 
transition to pairing: Two subsets are formed with antiparallel alignment 
between the subsets. This fixed point is probably unstable in the
$M\rightarrow \infty$ limit and only observed in simulations because
the self-averaging property of the ODEs breaks down at that point.
\end{itemize} 
Hence, with increasing learning rate the symmetry
of the system of interacting perceptrons is broken, but the state still
has the symmetry of the ring. Note that according to Eq. (\ref{Hebb-Eq}) 
the learning step scales to zero with system size $M$. 
The prefactor alone triggers the first phase transition.

Now we show that a system of interacting networks can show
better-than-random performance in a problem
of adaptive competition which was recently introduced by Arthur
\cite{Arthur} and  is being studied intensively 
\cite{Challet,Savit,Cavagna,Challet:Exact,Challet:Modeling}. 
It is a model of a closed market where 
$N$ agents are competing for
limited resources and where the individual profit depends on the action of
the whole community.

The model consists of $N$ agents who at each time step have to choose
between actions $\sigma^i = +1$ or $\sigma^i = -1, i = 1, \dots, N$. The
profit of each agent depends on the minority decision; each agent gains
$g^i = +1$ if he belongs to the minority, and he pays $+1$ if he belongs
to the majority of the common decision. Hence, one has 
$g^i = - \sigma^i
\mbox{sign} (\sum^N_{j=1} \; \sigma^j )$. 
The global
profit is given by 
\[ G = \sum\limits^N_{i=1} \; g^i = - \left|
\sum\limits^N_{j=1} \; \sigma^j \right| <0 \;\; , \] 
the cashier always
makes profit. Now each agent uses an algorithm which should maximize his
profit. In this model agents know only the history of the minority sign
$S_t = - \mbox{sign} ( \sum^N_{j= 1} \;\; \sigma_{t}^{j})$ 
for each previous time step $t$, and the agents are not allowed
to exchange informations.

If each agent makes a random decision $\sigma_i$, the mean square global
loss is 
\begin{equation} \label{sechs} \langle G^2 \rangle = N \;\; .
\end{equation} 
It is non trivial to find an algorithm which performs better
than (\ref{sechs}). Previous investigations studied algorithms where each
agent has two or more quenched random tables that prescribe decision 
for each of the $2^M$ possible histories
$\vec{x}_t = (S_{t-M+1}, \dots S_t)$. Each table receives a score, and
the one with the larger score is chosen. 

Here we introduce an approach where each agent uses the same dynamic
strategy.  We use a perceptron with adaptive weights for each agent to
make the decision.  The weights define the strategy of the agent, and
our strategies change with time as the weights are updated according to
the minority decision one time step earlier.  We follow the usual
scenario for training a perceptron: Start from a randomly chosen set of
initial weights and train each network by the usual Hebbian learning
rule.  At each time its decision of each agent is made by $\sigma_i =
\mbox{sign} (\vec{w}^i \cdot \vec{x})$ and each perceptron is trained by
the minority decision $S_t$,
\begin{equation} \label{sieben}
\vec{w}^i_{t+1} = \vec{w}_t^i - \frac{\eta}{M} \;
\vec{x}_t \mbox{sign} \left(\sum\limits^N_{j=1} \; \mbox{sign}
(\vec{x} \cdot \vec{w}^j) \right) \; \; 
. \end{equation}
Hence, the bit sequence $(S_t)$ is generated by the negative
output of a committee machine. From Eq. \ref{sieben} it follows 
that each weight vector is changed by the
same increment, hence only the center of mass of the weight vectors 
changes during the learning process.

Our numerical calculations show that starting from a random set of
weight vectors and initial input the systems relaxes to a state with a
good performance. The global gain is of the order of $N$ and for small
learning rates the system performs better than the random guessing.  We
succeeded to solve the dynamics of the interacting networks analytically
for the case where the input vector is replaced by a random one.

\begin{figure}
 \epsfxsize= 0.49\textwidth
 \epsffile{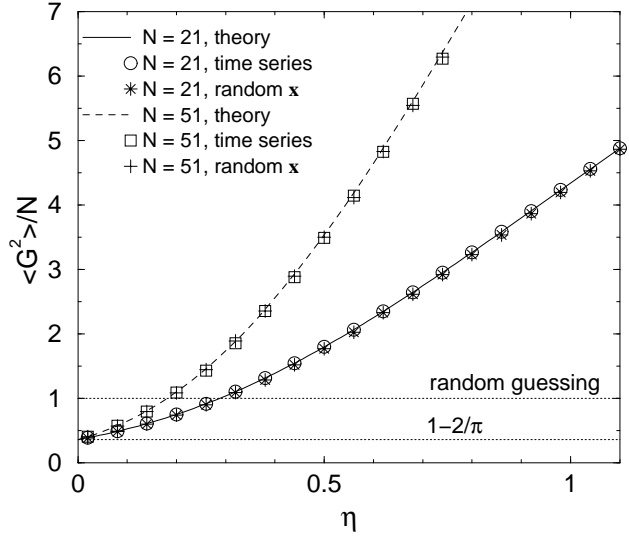} 
\caption{Average loss $\langle G^2 \rangle/N$ versus
learning rate in the Bar Problem, using learning
rule $H$. Simulations used $M=100$.}
\label{MIN-G_eta}
\end{figure}

Approximating the input $\vec{x}$ by a random one, we derived
the equation of motion of the norm of the center of mass;
the fixed point describes the global gain in the long run. 
To simplify the calculation, the initial norms
$|\vec{w}^i_0|$ are set to 1, the sum $\sum_i^N \vec{w}_0^i$
is 0, and the scalar products are symmetric: 
$\vec{w}^i_0 \cdot \vec{w}^j_0 =-1/(N-1)$ for $i\neq j$.
We obtain for the attractor of the dynamics  
\begin{eqnarray} \label{acht}
\langle G^2 \rangle/N &=& 
            1 + (N-1) \left( 1- \frac{2}{\pi} 
             \arccos \frac{A-1/(N-1)}{A + 1} \right);  \label{G-Fix} \\ 
A &=& \frac{\pi \eta^2}{16} 
     \left(1+ \sqrt{1 + \frac{16 (\pi -2)}{\pi N \eta^2}} \right) \; \; .
\label{Min-Fix} 
\end{eqnarray}
For random patterns, $A$ is the square norm of the center of mass
at the fixed point. Eq. (\ref{G-Fix}) 
agrees with simulations of both the real time series and
random patterns, as shown in Fig. \ref{MIN-G_eta}. Very similar results 
(up to factors of $1 + 1/\sqrt{N}$) are 
found analytically and in simulations by starting with uncorrelated
random vectors. For
small learning rate $\eta \rightarrow 0$ we obtain the best global gain 
\begin{equation}
\langle G^2 \rangle = \left(1 - \frac{2}{\pi}\right) N \simeq 0.363 N\; \;
.\label{Min-Limit} \end{equation}
It is interesting that this result is also obtained for a scenario where 
we use a distribution of learning rates $\eta$ instead of a fixed one. 
For every perceptron at every time step a different learning rate is
chosen. Hence the center of mass is not fixed during learning and the
weight vectors increase their lengths similar to a random walk. This
process decreases the average learning rate compared to the length and
leads to  the performance given in Eq. (\ref{Min-Limit}). 

 Hence, the system of interacting networks performs better than the
random decision. In fact, there are several advantages of the 
system of neural networks compared to the algorithm of scoring 
quenched random tables.

Firstly, 
the size $M$ of the history does  not have to be adjusted to the number 
of agents in order to perform better than random. 
Our analysis implicitly assumes that $N \le M$ and both 
$M$ and $N$ are large,
but simulations show good qualitative agreement even for $N=21$, $M=4$.
For small $M$, $\langle G^2 \rangle /N$ even tends to be smaller than
predicted for $M=\infty$. 
We suspect that a strong dependence on the ratio of players to
possible strategies only occurs when players have to 
pick from a set of fixed strategies, and is absent when 
they fine-tune one strategy. However, this point
still needs further investigation.

Secondly, on average all of the agents perform identically -- this
is also a consequence of the absence of quenched disorder. There is no
phase transition between a set of successful agents and losers, as found
in Ref. \cite{Savit} for the random tables. This is clear from the
geometrical interpretion: The center of mass does a random walk on a
hypersphere around the origin.  The radius depends on the learning rate;
if the radius is smaller than $\sqrt{N}$ (obtained from when adding up
$N$ random vectors of norm 1), the ``strategies'' are distributed better
than random. As the center of mass moves, each perceptron shifts from
the current majority side to the minority side and back.

Eq. (\ref{Min-Limit}) represents the optimum obtainable 
for perceptrons as long as the symmetry among them is
not broken. It would be
interesting to study other network architectures to see
whether the profit of a system of competing neural networks can still
be improved. 

This work benefitted from a seminar at the Max-Planck Institut f\"{u}r
Physik komplexer Systeme, Dresden. The authors want to thank Michael Biehl
and Georg Reents for useful discussions, and Andreas Engel and Johannes
Berg for their introduction to the minority game.

\end{document}